\documentclass[prb,superscriptaddress,showpacs,twocolumn,preprintnumbers,amsmath,amssymb,longbibliography]{revtex4-1}
\usepackage{graphicx}
\usepackage{color}
\usepackage{bm}
\usepackage{epstopdf}
\usepackage{times}
\usepackage{float}

\AtBeginDocument{%
    \newwrite\bibnotes
    \def\bibnotesext{Notes.bib}
    \immediate\openout\bibnotes=\jobname\bibnotesext
    \immediate\write\bibnotes{@CONTROL{REVTEX41Control}}
    \immediate\write\bibnotes{@CONTROL{%
    apsrev41Control,author="08",editor="1",pages="1",title="0",year="1"}}
     \if@filesw
     \immediate\write\@auxout{\string\citation{apsrev41Control}}%
    \fi
}%

\begin{document}

\title{Frozen-phonon method for state anticrossing situations and its application to zero-point motion effects in diamondoids}

\author{Pablo Garc\'ia Risue\~no}
\email[E-mail:]{garcia.risueno@gmail.com}
\affiliation{Institut f\"ur Physikalische Chemie, Universit\"at Hamburg, Grindelallee 117, D-20146 Hamburg, Germany}
\affiliation{The Hamburg Centre for Ultrafast Imaging, Luruper Chaussee 149, D-22761 Hamburg, Germany}
\author{Peng Han}
\affiliation{Institut f\"ur Physikalische Chemie, Universit\"at Hamburg, Grindelallee 117, D-20146 Hamburg, Germany}
\affiliation{Department of Physics, Capital Normal University, Beijing Key Lab for Metamaterials and Devices, Beijing 100048, China}
\author{Gabriel Bester}
\email[E-mail:]{gabriel.bester@uni-hamburg.de}
\affiliation{Institut f\"ur Physikalische Chemie, Universit\"at Hamburg, Grindelallee 117, D-20146 Hamburg, Germany}
\affiliation{The Hamburg Centre for Ultrafast Imaging, Luruper Chaussee 149, D-22761 Hamburg, Germany}

\date{\today}

\begin{abstract}
The frozen-phonon method, used to calculate electron-phonon coupling effects, requires calculations of the investigated structure using atomic coordinates displaced according to a certain phonon eigenmode. The process of ``freezing-in'' the specific phonon can bring electronic eigenstates that are energetically close in energy into an anticrossing. This electronic anticrossing effect is, however, unrelated to the wanted electron-phonon coupling, and needs to be removed. We present a procedure how to deal with these problematic anticrossing situations and apply it to the band gap zero-point motion renormalization of sixteen diamondoids and urotropine using different exchange correlation functionals. We find gap renormalizations of diamondoids in the range of 150 - 400 meV and only 62 meV for urotropine due to the lone-pair character of the highest occupied molecular orbital of the latter.
\end{abstract}


\maketitle

\section{Introduction}

Understanding the effect that the nuclear motion has on electronic properties has been a challenging scientific problem \cite{frohlich1954,holstein59a,*holstein59b,PhysRev.131.993,hedinep,AH76} and remains a central topic of solid state physics and quantum chemistry \cite{giustinoRMP,giustinoprl,ponceprl,marininatcomm,giustinoNC,PhysRevLett.101.106405}. Electron-phonon (e-ph) coupling can be studied with a computational technique referred to as ``frozen-phonon". As originally proposed by  Dacorogna {\it et al.}\cite{dacorogna85} to calculate the e-ph coupling matrix elements of bulk aluminum from first principle the idea is intuitive and simple: freeze the atoms into the displaced positions they acquire during a certain vibration. The electronic response to the distortion leads to the e-ph coupling. The method requires, however, large supercells especially for long wavelength phonons, which has limited its application. Furthermore, with the rapid development of the competing density functional perturbation theory (DFPT)\cite{savrasov94}, the frozen-phonon approach was rarely used to calculate e-ph coupling matrix elements\cite{han12b}. Later, Capaz {\it et al.}\cite{capaz05} extended the \emph{ab initio} frozen-phonon approach to study the renormalization of the band gap due to zero-point motion (ZPR), i.e. the modification of the electronic eigenvalues by a nuclear quantum effect, and the temperature-dependent band gap renormalization. Compared to the ``standard'' Allen-Heine-Cardona (AHC) theory of ZPR\cite{allen81}, the frozen-phonon approach includes the so-called non-diagonal Debye-Waller terms missing in the AHC model as well as some anharmonic effects of lattice vibration and gives a more accurate description on ZPR\cite{gonze2011,giustino17}. 
In recent years, the \emph{ab initio} based frozen-phonon approach has been widely used to calculate the ZPR effect and temperature-dependent band gap renormalization in semiconductor nanoclusters\cite{besterPRB2013}, in bulk diamond, Si, and SiC structures\cite{monserrat14}, in hexagonal and cubic ice\cite{engel15}, and in molecular crystals\cite{monserrat15,polaronpaper}. Moreover, recent developments of the frozen-phonon method include non-diagonal supercells,
 extending the applicability the approach\cite{williams15,monserrat16}, and one-shot calculation of temperature-dependent optical spectra \cite{zacharias16}.
However, the frozen-phonon method suffers, as we intend to demonstrate, from a problem that occurs when the applied deformation (frozen-phonon) leads to an anticrossing of the electronic states. This anticrossing effect overshadows the more subtle phononic effect on the electronic state.  

In this paper, we propose a modification of the standard frozen-phonon method\cite{capaz05} in order to properly treat the situation of state anticrossings, which can happen when eigenstates are energetically close to the eigenstate considered. 
We illustrate the problem of state mixing and show that four qualitatively different situations can occur. We present a method to circumvent the problems and apply it to sixteen carbon-caged structures called {\it diamondoids} and urotropine.
Our results show that the renormalization of the band gap due to electron-vibrational interaction in diamondoids is strong, in the same order of magnitude as in bulk diamond (about 600 meV). Moreover, we find that the results using hybrid functionals (B3LYP) are rather different from those of LDA and GGA for the renormalizations of the highest occupied molecular orbital (HOMO) and the lowest unoccupied molecular orbital (LUMO) individually, but similar for the renormalizations of the HOMO-LUMO gap. The results for the HOMO renormalization of urotropine is very different than for the diamondoids (significantly smaller), a fact that we explain by the lone-pair nature of the HOMO in urotropine.

\section{Theory}

\subsection{Frozen-phonon approach}\label{sec:tf}

The phonon theory is based on the solution of the {dynamical equation} \cite{ponceprb}:
\begin{equation}
\sum_{I \alpha} \ D_{I \alpha, J\alpha'} \  {\textrm{X}}^{\nu}_{ I \alpha} \ = \  \omega_{\nu}^2 \  {\textrm{X}}^{\nu}_{J \alpha'} \ ,
\end{equation}
where $I,J=1,\ldots,N$ 
are atomic indices, $\alpha,\alpha'=1,2,3$ are their corresponding Cartesian coordinate indices, $\nu$ is the phononic index, and $ {\textrm{X}}^{\nu} $ (which are vectors of $3N$ components) are the normal modes of vibration. The dynamical matrix $D$ is defined as
\begin{equation}
D_{I \alpha, J \alpha'} \  \equiv \ \frac{1}{\sqrt{M_I M_J}} \ \frac{\partial^2  \mathcal{E}^{BO}}{ \partial  R_{I \alpha}  \partial  R_{J\alpha'}   }     \ ,
\end{equation}
where $M_I$ are the atomic masses, $ R_{I \alpha} $ are the nuclear coordinates and $\mathcal{E}^{BO}$ is the Born-Oppenheimer energy surface of the system \cite{ponceprb};
$\omega_{\nu}^2$ and ${\textrm{X}}^{\nu}$ are the eigenvalues and eigenvectors of the dynamical matrix ($\omega_{\nu}$ are the phonon frequencies).

The frozen-phonon method \cite{capaz05} is used to calculate the temperature-dependent renormalization of electronic eigenvalues $E_n$ due to vibronic coupling. Such renormalization is given by:
 \begin{equation}\label{eqrenorm1}
\Delta  E_n (T) \ = \ \sum_{\nu} \,   \Delta  E^{\nu}_n (0)  \  \left(  n^{\textrm{B}}_{\nu}  +  \frac{1}{2} \right) \ ,
\end{equation}
where $n^{\textrm{B}}$ is the Bose-Einstein distribution, in atomic units: ${n^{\textrm{B}}_{\nu} = ( \textrm{exp}[ \omega_{\nu}/T ] -1 )^{-1} }$. 
The $ \Delta E^{\nu}_n (0)  $ coefficients of (\ref{eqrenorm1}) are:
\begin{equation}\label{eqrenorm2}
\Delta E^{\nu}_n (0)   \, = \, \frac{1}{2 \omega_{\nu}} \,  \frac{d^2}{d h^2}  \, {E}_n  \left. [ \, {\bf x}_0 + h {\bf U}^{\nu} \, ]  \right|_{h=0} \, ,
\end{equation}
where $h$ is a displacement parameter (with units of $1/\sqrt{\omega_{\nu}}$),
$U^{\nu}_{I\alpha} = X^{\nu}_{I\alpha}  / \sqrt{m_{I}}$, $m_I$ the mass of the $I$-th atom,
 and ${E}_n[\bf x]$ is the $n$-th electronic eigenvalue at $T$ = 0 when the nuclei are at positions given by $\bf x$ (which is a vector of 3$N$ components),
 and ${\bf x}_0$ is the set of relaxed nuclear positions.
Eqs.~(\ref{eqrenorm1}) and (\ref{eqrenorm2}) result from performing an average of the electronic eigenvalues assuming a parabolic dependence on the nuclear positions. A derivation of these equations can be found in Ref. [\onlinecite{ponceprb}]. In the frozen-phonon method, the second  derivatives of (\ref{eqrenorm2}) are calculated by finite-difference. In the standard definition of the frozen-phonon method \cite{capaz05,cohenff,giustinoNC,FPalaCapazusage2,giustinoJPCM,besterPRB2013,FPalaCapazusage1,bester2016}, the displaced positions used in the finite difference are $ {\bf x}_\pm = {\bf x}_0 \pm {\bf U}^{\nu} / \sqrt{\omega_{\nu} }$. This results into the simple form:
\begin{equation}\label{eqcoefscapaz}
\Delta E^{\nu}_n (0)   =
\frac{
E_n [ {\bf x}_+ ]  - 2 E_n [ {\bf x}_0 ]  \, + \, E_n [  {\bf x}_-  ]
}{2} \quad .
\end{equation}

\subsection{Problematic situations in the case of anticrossings}\label{secproblematic}

\begin{figure*}
\includegraphics[width=\textwidth]{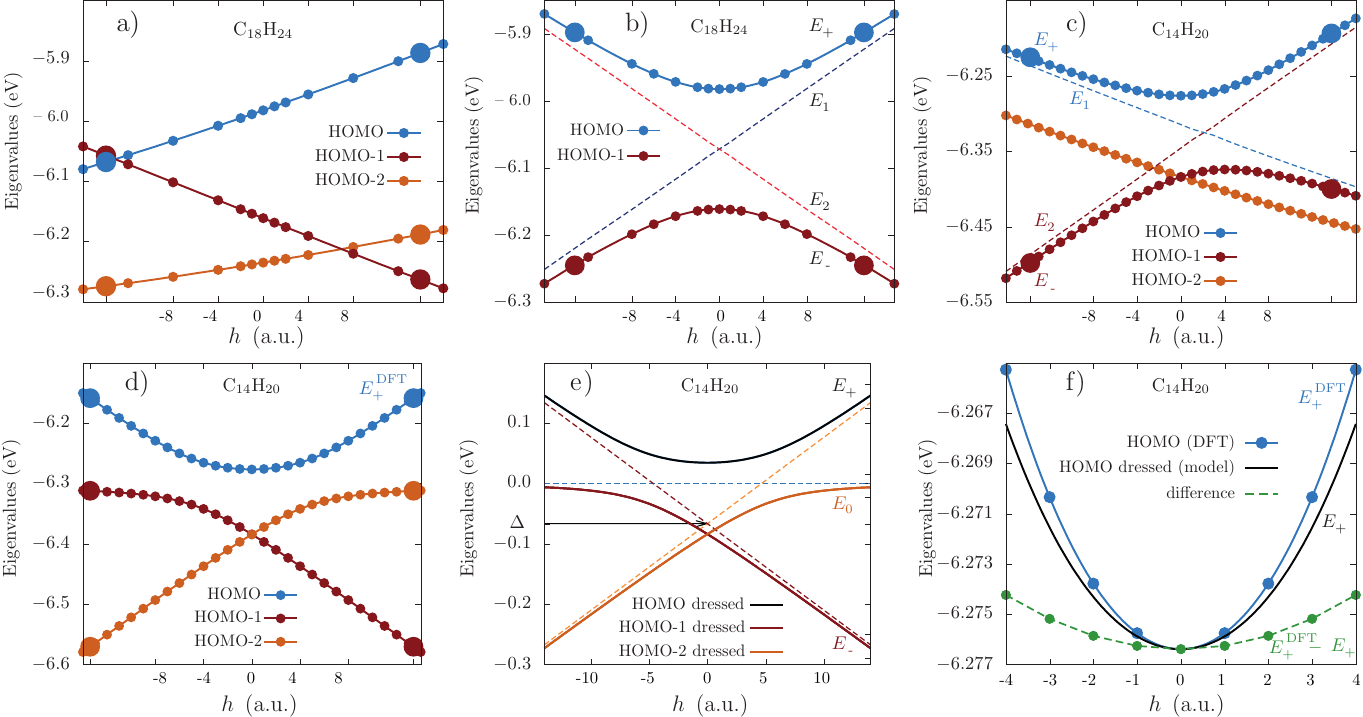}
\caption{Eigenvalues as a function of the frozen-phonon displacements $h$ for
a) C$_{18}$H$_{24}$ for a phonon with wavenumber 1131 cm$^{-1}$; 
b) C$_{18}$H$_{24}$, for a phonon with wavenumber 1256 cm$^{-1}$; 
c) C$_{14}$H$_{20}$, for a phonon with wavenumber 1217 cm$^{-1}$.  
d) C$_{14}$H$_{20}$, for a phonon with wavenumber 1301 cm$^{-1}$; 
e) bare and dressed states extracted from panel (c) (see text);
f) HOMO from DFT (as in panel (d)) and HOMO from the model (as in panel (e)) and difference between both (dashed).}
\label{Fig.mc3}
\end{figure*}

In Fig. \ref{Fig.mc3} we show the highest occupied eigenvalues of lower diamondoids as a function of the displacement size $h$ (with $ h  {\bf U}^{\nu}$ displacements) for different vibrational modes. The large dots indicate displacements according to the original work \cite{capaz05} (i.e. with  ${  {\bf U}^{\nu} / \sqrt{\omega_{\nu} }}$ displacements). In Fig.~\ref{Fig.mc3} the different problematic situations are shown, corresponding to  (a) a simple level crossing, (b) a symmetric anticrossing (avoided crossing) of two levels (c) a combined situation of crossing and anticrossing and (d) a double anticrossing. The anticrossing situations (b,c,d) are problematic because the curvatures of the $h$-dependence of the eigenvalues have a dominant component coming from the anticrossing effect itself and not from the ZPR. In Fig.~\ref{Fig.mc3}a), a more trivial problem can occur, when the states simply cross and one must take care to follow the HOMO or LUMO across the crossing and not simply take the HOMO and LUMO states in the distorted structure.

\subsection{Circumventing the problem of state crossings}\label{sec:amff}

We propose to use the following equation, instead of Eq.~(\ref{eqcoefscapaz}):

\begin{eqnarray}\label{eqcoefs1}
\Delta E^{\nu}_n (0)   =  \frac{    E_n [  {\bf x}_0 + h  {\bf U}^{\nu} ] - 2 E_n [  {\bf x}_0] + E_n [  {\bf x}_0  - h  {\bf U}^{\nu}] }{2 \ \omega_{\nu} \  h^2 }  \, \, \, \, \, \,  \, \, \, \,
\end{eqnarray}
with $h$ 
as small as possible \cite{notedelta2}, e.g. between 2 and 10 a.u..
The use of a small $h$ circumvents the problem of the eigenvalue crossings shown in Fig.~\ref{Fig.mc3},a) in all the cases we have experienced (see Supporting Information).

\subsection{Corrections necessary for the problem of state anticrossings}\label{sec:amff}

In order to recognize the situations (i.e., the vibrational modes) where a state anticrossing (or crossing) takes place we propose the following procedure.
\begin{itemize}
\item[1)] Solve the dynamical equation (see Sec. \ref{sec:tf})
\item[2)] Calculate the electronic eigenvalues ($E_n$) at the relaxed (undisplaced) position ${\bf x}_0$, and also at the displaced positions ${\bf x}_0 + h  {\bf U}^{\nu}$ and ${\bf x}_0 - h  {\bf U}^{\nu}$  for all vibrations $\nu$ using a small $h$ ($\approx$ 2  a.u.).
\item[3)] Calculate the overlap between undisplaced $| u \rangle$ and displaced $| d \rangle$ states $\chi = | \langle u | d \rangle|^2$.
If the overlap is close to 1.0, calculate the renormalization according to Eq. (\ref{eqcoefs1}). If the overlap is less than 1.0 (a value of 0.995 for $h$ = 2, was used in this work), proceed with 4) and 5)
\item[4)] Recalculate the eigenvalues for a full range of displacements  $h = \pm 1, \ \pm 3, \ \pm 4, \ldots $
\item[5)] By visual inspection, identify the anticrossing situation (Fig.~\ref{Fig.mc3} b),c) or d)) and correct accordingly, as described next.
\end{itemize}

\noindent{\underline{Avoided crossing of two states, Fig.~\ref{Fig.mc3} b) or  \ref{Fig.mc3} c) }}\\

We map the simple state anticrossing situation (Fig.~\ref{Fig.mc3},b) as well as the asymmetric anticrossing in Fig.~\ref{Fig.mc3},c) to the equation:
\begin{equation}\label{eqanticrossd}
\left|   \, \, \,
\left( \,
\begin{array}{cc}
 E_1  &      g  \\
             g         & E_2
\end{array}
\right)
\,  - \mathbb{I}  \cdot \, E_{\pm}  \,
\,
\right| \, = \,0 \ ,
\end{equation}
where $\mathbb{I} $ is the identity matrix $E_{\pm}$ are the eigenvalues we obtain from DFT  and $E_{1,2}$ are the unknowns, the so-called bare states. The $E_{1,2}$ energies do not include the anticrossing effect, but do include the ZPR. We therefore extract the ZPR effect from $E_{1,2}$. Eq. (\ref{eqanticrossd}) has an analytic solution:
\begin{equation}
\label{eqanticrossd2}
E_{1\atop 2} \,   = \frac{  E_{+} + E_{-}   \, \pm \,  \sqrt{  (E_{+} - E_{-})^2 - 4g^2 }  }{2} \quad .
\end{equation}
The coupling parameter $g$ is half the minimal distance between the dressed curves ($E_{+}$ and $E_{-}$) and is obtain from a polynomial fit to the DFT eigenvalues (in the case of Fig.~\ref{Fig.mc3},(b) it is simply  $E_{+} - E_{-}$ taken at $h = 0$).

The procedure is as follows:
\begin{itemize}
\item[1) ] Determine the coupling parameter $g$ from the minimum distance.
\item[2) ] For each $h$, $E_{+}(h)$, $E_{-}(h)$, solve Eq. (\ref{eqanticrossd2}).
\item[3) ] Fit the $E_{1,2}$-vs-$h$ to a parabola, and extract its second derivative (curvature) $\zeta$.
\item[4) ] Calculate the frozen-phonon renormalization using  $\Delta E_n^\nu(0) = \zeta / (2 \omega_{\nu})$.
\end{itemize}

In Fig.~\ref{Fig.mc3},b) and  \ref{Fig.mc3},c) we show the calculated bare eigenvalues (dashed lines) as well as the dressed eigenvalues from DFT (circles). The solid lines are polynomial fits to the data points. \\

\noindent{\underline{Avoided crossing of 3 states, Fig.~\ref{Fig.mc3},d)}}\\

The Hamiltonian describing the anticrossing situation in Fig.~\ref{Fig.mc3},d) is given by:
\begin{equation}
\label{eqanticrosst}
\hat{\rm H} =
\left( \,
\begin{array}{ccc}
 \Delta + a h &      g & g_3  \\
   g         & 0  & g \\
   g_3  & g  & \Delta - a h
\end{array}
   \right) \quad ,
\end{equation}
with the solution shown in Fig.~\ref{Fig.mc3},e).
The eigenvalues at $h = 0$ are given by:
\begin{subequations}
\begin{align}
E_+ (h=0) &= \frac{1}{2} \left( \, \Delta + g_3 + \sqrt{8g^2 + (\Delta + g_3)^2} \, \right) \\
E_0 (h=0)&= \frac{1}{2} \left( \, \Delta + g_3 - \sqrt{8g^2 + (\Delta + g_3)^2}  \, \right)\\
E_- (h=0)&= \Delta - g_3 \, .
\end{align}
\end{subequations}

The requirement of a crossing of the states $E_-$ and $E_0$ at $h = 0$, which is dictated by the symmetry, leads to an analytic solution for the coupling $g_3$:
\begin{equation}
\label{eq:g3}
g_3 = \frac{1}{2} (\Delta + \sqrt{\Delta^2 + 4g^2}) \quad .
\end{equation}

The procedure for the situation shown in Fig.~\ref{Fig.mc3},d) is as follows.

\begin{itemize}
\item[1)] Determine $a$ and $\Delta$ by using the large $h$ limit of the ab-initio calculated results.
\item[2)] Calculate $g$  as the minimal distance between HOMO and HOMO-1 using the DFT results.
\item[3)] Calculate $g_3$ using Eq.(\ref{eq:g3}).
\item[4)] Solve $\hat{\rm H}$ for each $h$ value to obtain the dressed eigenvalues from the anticrossing model: $E_{+,0,-}$.
\item[5)] Calculate the difference between the eigenvalues from DFT $E_{+}^{\rm DFT}$ and the dressed eigenvalues from the model
$E_{+}$ for every $h$.
\item[6)] Extract the concavity of the curve obtained in 5),  $\zeta$.
\item[7)] Calculate the frozen-phonon renormalization using $\Delta E_n^\nu(0) = \zeta / (2 \omega_{\nu})$.
\end{itemize}

\begin{figure}
\includegraphics[width=\columnwidth]{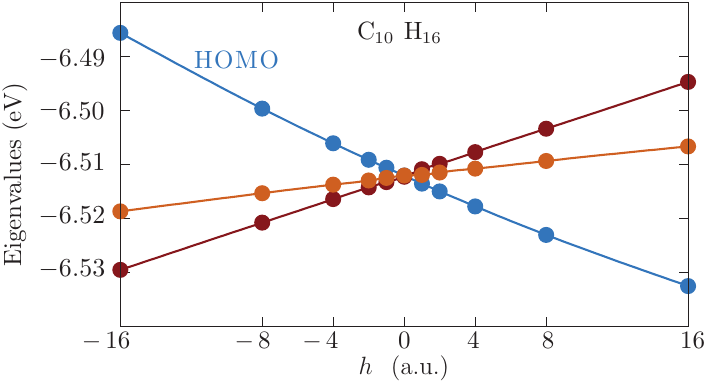}
\caption{Triply degenerate HOMO of adamantane as a function of the displacement parameter $h$ for a phonon with wavenumber 740 cm$^{-1}$.}
\label{Fig.degeneracy}
\end{figure}

In Fig.~\ref{Fig.mc3},f) the dressed eigenvalues of the anticrossing model $E_{+}$ is shown as solid black lines and the DFT eigenvalues as blue circles $E_{+}^{\rm DFT}$. The difference between both is shown as dashed green line. From this difference we extract the second derivative in step 6) above.
In the analyzed example, the uncorrected (corrected) renormalizations is 56.2 (11.8) meV. \\

\noindent{\underline{Treatment of degenerate states}}\\

For degenerate states, as shown in Fig. \ref{Fig.degeneracy} (HOMO of adamantane), we average the renormalization of the degenerate states.

 \section{Structures investigated}

\begin{figure*}
\includegraphics[width=5.05in]{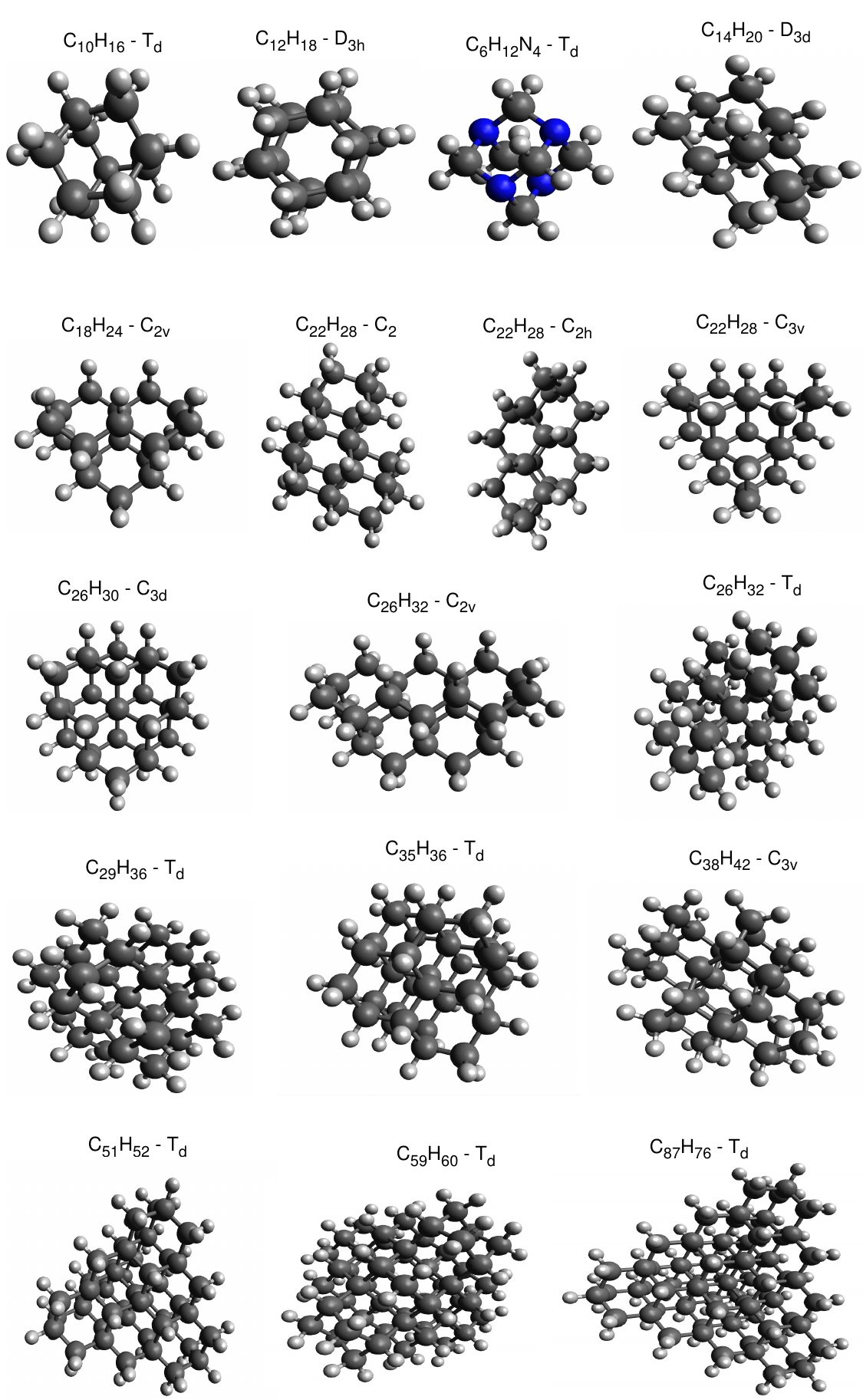}
\caption{Nanostructures investigated in this article (16 diamondoids and urotropine).}
\label{Fig.diamondoids}
\end{figure*}

We investigated sixteen diamondoids and urotropine (also known as hexamethylenetetramine, C$_{6}$H$_{12}$N$_{4}$), where four methine groups of adamantane are replaced by nitrogen, as shown in Fig. \ref{Fig.diamondoids}. 
Diamondoids are ideal candidates to study ZPR. These small fragments of diamond passivated with hydrogen present a strong ZPR \cite{marininatcomm,giustinoNC,bester2016} and can be simulated {\it ab initio} with an affordable computational burden.
Diamondoids present many forms, like those of a cube, a cane, a disk or a pyramid \cite{reviewdiamondoids3}.
Three of the smallest diamondoids, adamantane (C$_{10}$H$_{16}$), diamantane (C$_{14}$H$_{20}$) and triamantane (C$_{18}$H$_{24}$), usually called \emph{lower diamondoids}, have no isomers, in contrast to larger \emph{higher diamondoids}.
The systems that we analyze in this article have been object of \emph{ab initio} calculations in the past to different extent. On the one hand, the lower diamondoids are object of an intense computational \cite{giustinoNC,marininatcomm} and experimental \cite{diamondoidexperimental2,diamondoidexperimental3} research. Higher diamondoids, on the other hand, have been object of fewer \cite{sintesisdiamondoids1}  experiments and several computational studies  \cite{paperC87,paperc87-2,diamondoidsinspace,diamondoidsinspace2,diamondoidsinspace3,paperc87-2,paperc29h36-3,paperc29h36-3,paperc59h60,paperc59h60-2}. \\

\section{Computational methods}

We performed our DFT calculations using the \texttt{Quantum Espresso} (v.5.3.0) package \cite{QEpaper}. The exchange-correlation functionals were LDA/Perdew-Zunger parametrization \cite{paperldapz} with the VBC pseudopotential \cite{papervbc}, GGA/PBE \cite{paperpbe} version with the HGH pseudopotential \cite{paperhgh}, and B3LYP \cite{Bec1993JCP,Lee1988PRB}. In the calculations with B3LYP we used the relaxed geometries and normal modes calculated with GGA-PBE.
The plane-wave cutoffs were 30 Ry for GGA and 80 Ry for LDA (except for C$_{22}$H$_{28}-C_2$, 30 Ry) and 60 Ry for B3LYP.
The simulation cells were cubic, with edges of 35 \AA {} for systems up to 18 carbons, 36 \AA {} for
systems with between 22 and 35 carbons,
37 \AA {} for systems with 38 and 51 carbons and 38 \AA {} for the largest diamondoid
(we used very large simulation cells to accurately capture the surface bound states of nanodiamonds \cite{bester2016}). The forces were converged to $ 10^{-6}$ Ry/Bohr.

\section{Results}\label{sec:results}

Our results for the renormalization of the HOMO, LUMO and gaps are summarized in Table \ref{Tab.results}, where we also present the renormalization of the LUMO+1 of C$_{14}$H$_{20}$ and C$_{22}$H$_{28} - C_{2h}$, because the HOMO/LUMO transition is forbidden due to symmetry arguments \cite{bester2016}.

\begin{table}
\caption{ZPR of the HOMO and LUMO states of diamondoids and urotropine in meV. The numbers in bold font indicate the number of modes where a correction due to an anticrossing was necessary.}
\label{Tab.results}
\begin{tabular}{cccccc}
\hline
\hline
	~~~~ System ~~~~~& ~~~~~State~~~~~   & ~~~~GGA ~~~~~ & ~~~~~LDA~~~~~ & ~~~~ B3LYP~~~~\\
	~~~~ (symmetry) ~~~~~& ~~~~~ ~~~~~   & ~~~~  ~~~~~ & ~~~~~ ~~~~ & ~~~~ ~~~~\\
\hline
\hline
C$_{ 10}$H$_{16 }$     &  HOMO     &  162.4        & 158.3         & 202.9  \\
  ($T_d$)              &  LUMO     &  -86.6        & -95.1         & -71.4  \\
\hline
C$_{ 12}$H$_{18 }$    &  HOMO     &  224.2        &  197.3         & 295.7  \\
  ($D_{3h}$)                    &  LUMO     &  -92.2        & -113.5         & -73.8  \\
\hline
C$_{ 14}$H$_{30 }$     &  HOMO     &  268.9 {\bf  8}        &  317.7 {\bf  4}    & 314.4 {\bf  8}  \\
  ($D_{3d}$)       &  LUMO     &  -91.3                  & -105.5         & -49.3  \\
~~~~~~~~~~~~~~~~~~~~~~~~       &  LUMO+1     &  -67.9        & -72.0         & -23.7  \\
\hline
C$_{ 18}$H$_{24 }$    &  HOMO     &  183.2 {\bf 2}       &  211.0 {\bf 2}       & 239.2 {\bf 4}  \\
   ($C_{2v}$)       &  LUMO     &  -84.2               & -97.1         & -57.2  \\
\hline
C$_{ 22}$H$_{28 }$     &  HOMO     &         &  234.2 {\bf 18}       & \\
   ($C_{2}$)       &  LUMO     &                     & -106.9         &   \\
\hline
C$_{ 22}$H$_{28 }$    &  HOMO     &  177.5 {\bf 1}       &  160.3 {\bf 1}       & 239.2 {\bf 2}  \\
   ($C_{2h}$)       &  LUMO     &  -66.5               & -98.0         & -44.6  \\
~~~~~~~~~~~~~~~~~~~~~~~~       &  LUMO+1     &  -80.7               & -99.9         & -83.6  \\
\hline
C$_{ 22}$H$_{28 }$     &  HOMO     &  188.8        &  182.6       &   240.6 \\
  ($C_{3v}$)       &  LUMO      &  -85.5        & -95.1   &   -58.8  \\
\hline
C$_{ 26}$H$_{30 }$     &  HOMO     &  167.2        &  156.7       &   221.2 \\
  ($C_{3d}$)       &  LUMO      &  -73.9        & -99.3    &   -46.2  \\
\hline
C$_{ 26}$H$_{32 }$     &  HOMO     &  200.4        &  188.5       &   266.7 \\
  ($C_{2v}$)       &  LUMO      &  -80.2        & -89.1   &   -59.4  \\
\hline
C$_{ 26}$H$_{32 }$     &  HOMO     &  136.4        &  126.8       &   170.8 \\
  ($T_{d}$)       &  LUMO      &  -73.7        & -105.3   &   -70.3  \\
\hline
C$_{ 29}$H$_{36 }$     &  HOMO     &  175.2        &         &   187.2 \\
  ($T_{d}$)       &  LUMO      &  -93.9        &    &   -72.4  \\
\hline
C$_{ 35}$H$_{36 }$     &  HOMO     &  127.5        &  124.9       &   168.7 \\
 ($T_{d}$)       &  LUMO      &  -95.5        & -111.3   &   -81.5  \\
\hline
C$_{ 38}$H$_{42 }$    &  HOMO     &  205.7 {\bf 4}        &  243.6 {\bf 4}       &   255.1 {\bf 4} \\ 
	($C_{3v}$)       &  LUMO      &  -117.9        & -110.6   &   -75.5  \\
\hline
C$_{ 51}$H$_{52 }$     &  HOMO     &  101.9        &  110.1       &   \\
 ($T_{d}$)       &  LUMO      &  -100.6        & -82.3   &     \\
\hline
C$_{ 59}$H$_{60 }$     &  HOMO     &  192.8        &         &    \\
 ($T_{d}$)       &  LUMO      &  -101.1        &    &     \\
\hline
C$_{ 87}$H$_{76 }$     &  HOMO     &  96.6        &  91.9       &    \\
  ($T_{d}$)       &  LUMO      &  -61.3        & - 79.8  &     \\
\hline
\hline
C$_{ 6}$H$_{12 }$N$_{4}$     &  HOMO     &  -29.5        &  -36.3       &   -11.9 \\
 ($T_{d}$)                               &  LUMO      &  -68.7        & -76.6   &   -50.3  \\
\hline
\hline
\end{tabular}
\end{table}

\begin{figure}
\includegraphics[width=\columnwidth]{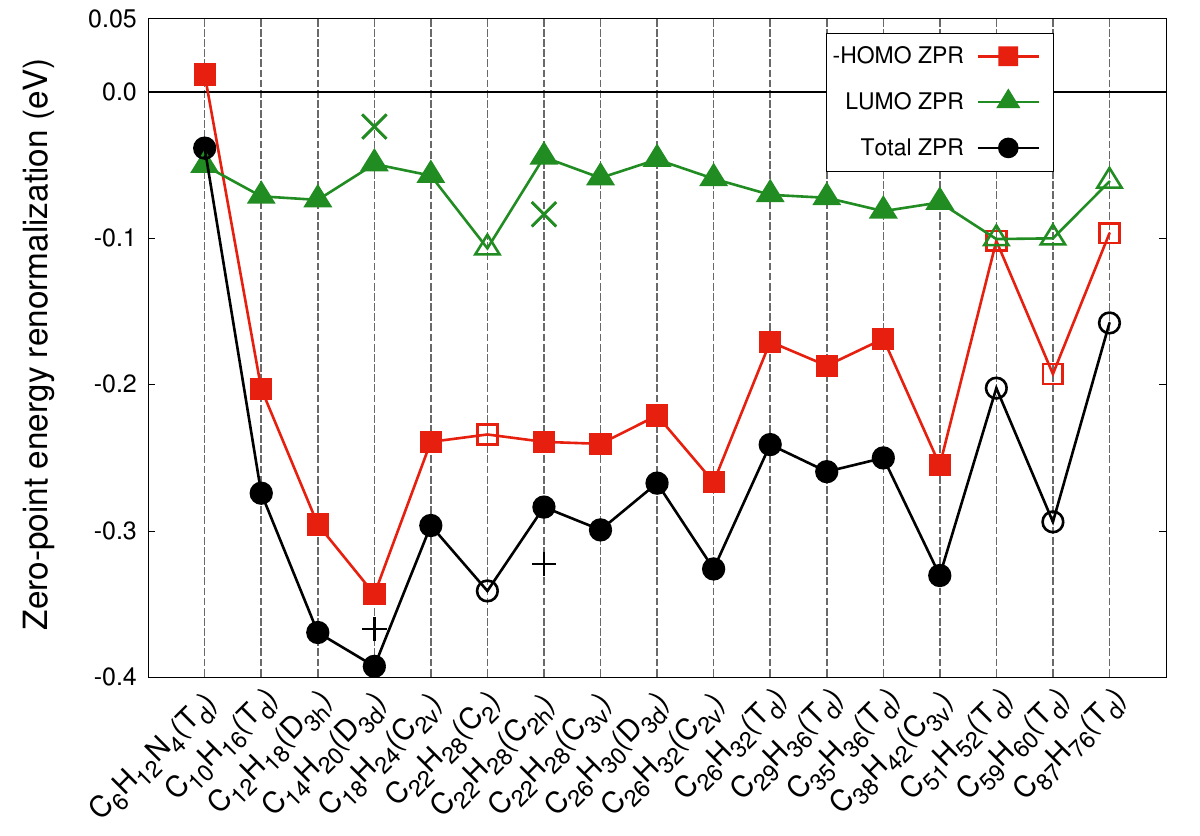}
\caption{\label{Fig.results} {
Zero-point renormalizations of urotropine and diamondoids. The HOMO renormalization is shown with opposite sign. Solid symbols for B3LYP functional, hollow symbols for GGA-PBE functional (except C$_{22}$H$_{28}$-$C_2$, using LDA-PZ). The $\times$ sign indicates renormalization of the LUMO+1, and the $+$ sign is the HOMO/LUMO+1 gap.
}}
\end{figure}

\begin{figure}
\includegraphics[width=\columnwidth]{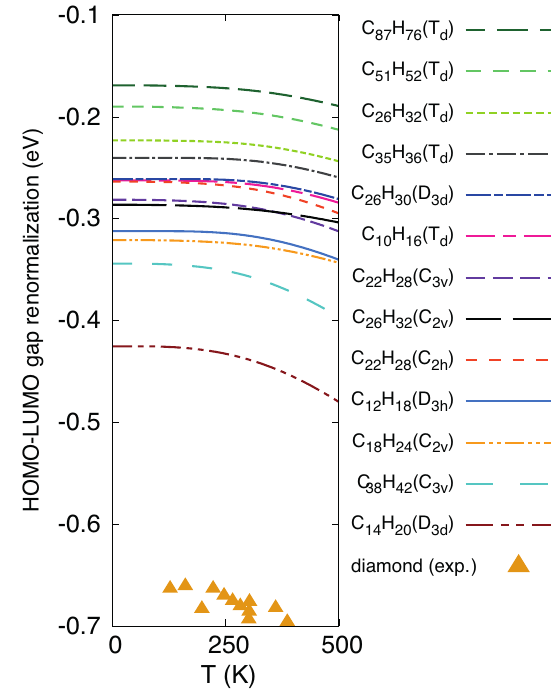}
\caption{\label{Fig.results4} {
Temperature dependence of the HOMO-LUMO gap (LDA), neglecting lattice expansion effects.
 }}
\end{figure}

In figure \ref{Fig.results} we present the frozen-phonon renormalizations of HOMO, LUMO and HOMO-LUMO gap of the 16 analyzed diamondoids and urotropine calculated with B3LYP, LDA and GGA functionals.
We notice that the renormalizations of the HOMO are always positive (for diamondoids, not including urotropine) and the renormalizations of the LUMO are always negative, which is the usual behavior \cite{ponceprl,besterPRB2013}. In addition, we notice that the magnitude of the HOMO renormalization is always larger than that of the LUMO renormalization, and that the values of the LUMO renormalizations are very similar for all the analyzed diamondoids (about -100 meV for LDA and GGA functionals, and about -70 meV for B3LYP). The latter property is attributed to the fact that the LUMO states are localized just outside the diamondoids and have surface-bound character due to the negative electron affinity \cite{bester2016}.
The values of the gap renormalizations lie between -158 meV (C$_{87}$H$_{76}$) and -393 meV (C$_{14}$H$_{20}$).
Note that for the LUMO renormalization of C$_{60}$ LDA makes an error of about 30\%, while the error of B3LYP is only 10\% \cite{c60scissor}(taking self-consistent $GW$ calculations as the reference).
Our results show an average differences between LDA and B3LYP  of -30\% for the HOMO, 35\% for the LUMO, and only 8.6\% for the HOMO-LUMO gap due to a cancellation of errors.

From the obtained results we conclude that:
\begin{itemize}
\item The correction of anticrossing effects can have a large effect on the calculated ZPR. The most striking difference is for diamantane (C$_{14}$H$_{20}$): for B3LYP, the corrected/uncorrected renormalization of the HOMO is (341 meV/562 meV. A similar reduction is obtained for triamantane (C$_{18}$H$_{24}$) with a corrected/uncorrected renormalization of 239 meV/ 446 meV.
\item The total ZPR renormalizations of diamondoids are of the order of 150 - 400 meV, which is the same order of magnitude as the renormalization of bulk diamond (which is about -628 meV \cite{ponceprl}). Furthermore, the ZPRs of diamondoids are smaller compared to the case in nanodiamonds (larger carbon clusters with diameters of 1.3 to 1.6~nm) where a ZPR between 0.85 and 0.9 ~eV was obtained\cite{besterPRB2013}.
\item The ZPR renormalization of the LUMO is similar, and rather small, for all the analyzed diamondoids; in agreement with the fact that the LUMO are surface bound states\cite{bester2016} which are localized mainly just outside the structures. The renormalization of the HOMO is size dependent with a general trend to decrease with increasing size (similar to the case of Si nanoclusters \cite{besterPRB2013}).
\item The renormalizations are very different for urotropine. We can understand this difference from the electronic structure of the HOMO. In urotropine, the HOMO is constituted of the lone pairs and is very different and much more delocalized than in the case of adamantane, where the HOMO is well localized inside the molecule (see Supporting Information). The ZPR in urotropine is therefore similar for HOMO (lone pair states\cite{banerjee15}) and LUMO (surface bound states \cite{bester2016}).
\item The temperature dependence shown in Fig. \ref{Fig.results4} is strongest in C$_{14}$H$_{20}$ and in  C$_{38}$H$_{42}$, but it hardly reaches 30 meV between 0 and 400 K, in agreement with previous work \cite{giustinoNC}. Note that our calculations do not include changes in the bond lengths (this anharmonic effects has a minor contribution to the renormalization in other carbon-based materials, like diamond \cite{PhysRevB.45.3376} and nanotubes \cite{capaz05}).
\item The results provided by GGA and LDA are very similar, as shown in Fig. \ref{Fig.results}. B3LYP provides higher values for the HOMO and LUMO renormalizations compared to LDA/GGA, but similar results for the HOMO-LUMO gaps.
\end{itemize}

\section{Conclusion}

We have introduced a method to perform accurate frozen-phonon calculations of the ZPR in the case where state anticrossing are influencing the results, i.e. when states are energetically close to the HOMO or LUMO. The new procedure is certainly more cumbersome than the original \cite{capaz05} procedure, since it requires a wavefunction projection to identify the problematic cases and a subsequent case-by-case correction on the specific modes. However, since it seems that only relatively few modes (at least for the investigated structures) require post-processing, the method remains reasonable and competitive.
The application of our method to the ZPR of 16 diamondoids and one modified diamondoid --urotropine-- confirms a rather strong gap renormalizations for the diamondoids and predicts a small renormalization for urotropine. 
The LDA and GGA results for the band gap renormalizations lie close to those obtained with B3LYP although the HOMO/LUMO results differ by up to 35\%.

\indent{ }

\noindent{\bf \large \center {$\qquad $ $\qquad $ $\qquad $ $\qquad $$\qquad $ $\qquad $  $\qquad $ $\qquad $ $\qquad $ $\qquad $$\qquad $ $\qquad $ } Acknowledgments {$\qquad $ $\qquad $ $\qquad $ $\qquad $$\qquad $ $\qquad $ } \\}
\indent{}
Most computations were performed on the HPC cluster of the Regional Computing Center of the Universit\"at Hamburg.
PGR like to thank Neil Drummond and Farah Marsusi for providing the structure of C$_{87}$H$_{76}$ and to Dr. Hinnerk St\"uben for valuable help and advice.


%

\newpage

\begin{widetext}

\begin{center}
{\bf \LARGE Supporting Information}
\end{center}

\section*{Choice of displacement parameter $h$ and overlap criterium}\label{sec:pp}

In the standard definition of the frozen-phonon method \cite{capaz05}, finite-difference calculations are performed on displaced positions ${\bf x}_0 \pm {\bf U}^{\nu} / \sqrt{\omega_{\nu} }$. This gives:
  \begin{equation}\label{eqcoefscapaz}
\Delta  E^{\nu}_n (0)   =  \frac{1}{2} \, \left( \,   E_n [ {\bf x}_0 +  {\bf U}^{\nu} / \sqrt{\omega_{\nu}} ] \, - \, 2 E_n [ {\bf x}_0 ]  \, + \, E_n [  {\bf x}_0  -  {\bf U}^{\nu} / \sqrt{\omega_{\nu}}   \, ]   \, \right)   \, .
\end{equation}
 The frozen-phonon method with such standard definition has been used in many recent articles \cite{giustinoNC,FPalaCapazusage2,giustinoJPCM,besterPRB2013,FPalaCapazusage1,bester2016}, and it has been successfully used to give account of the renormalization of electronic properties due to nuclear vibrations. As stated in our main article, this method is efficient and easy to implement, yet it presents one potentially important disadvantage: the sizes of the displacements (i.e. the norm of the ${\bf U}^{\nu} / \sqrt{\omega_{\nu}} $ vectors) are often too large, which can result into a \emph{mixing of states}  which distorts electronic eigenvalues and eigenvectors.   For most of the vibrational modes, the squared overlap between the electronic wavefunctions calculated at undisplaced and displaced nuclear positions $| \langle \textrm{u}  |  \textrm{d}  \rangle |^2 $ is high enough. That means the displaced and undisplaced wavefunctions essentially correspond to the same quantum state.   However, in some cases this does not hold, which
distorts the result of the renormalization and makes it unreliable.
For example in C$_{22}$H$_{28}$ with $C_2$ point-group symmetry, the displacements performed as presented in Ref. [\onlinecite{capaz05}] lead to mode 58  (out of 261 modes) with $ \chi  \equiv | \langle \textrm{u}  |  \textrm{d}  \rangle |^2 < 0.95$, to mode 46  with $ \chi  < 0.90$ and to mode 16  with  $ \chi  < 0.80$.
The displacements that we propose (i.e. ${ \pm h  {\bf U}^{\nu} }$ instead of ${ \pm {\bf U}^{\nu} / \sqrt{\omega_{\nu} }}$)
allow to use a much smaller perturbation and hence avoid some of the problems. A value of $1/\sqrt{\omega_{\nu}}$, as we will see, is often too large. In addition, the fact that this value depends on the phonon frequency makes it vibrational mode dependent. This makes the size of the displacements somewhat inconsistent. Hence, we remove the $1/\sqrt{\omega_{\nu}}$ term from the expression of the displacement vectors. Instead, a parameter $h$ is used to describe the amplitude of the lattice displacement. 

We have calculated the frozen-phonon renormalization of the HOMO and LUMO of diamondoids by using displacements as presented in Ref. [\onlinecite{capaz05}] and also using the expression that we propose (${ \pm h  {\bf U}^{\nu} }$, for different values of $h$). We found large differences (30\%) in the renormalization of the HOMO of triamantane (C$_{18}$H$_{24}$; 221 vs 288 meV);
 in the renormalization of the the HOMO of C$_{14}$H$_{20}$  the difference is 38\% (261 vs 359 meV);
 in C$_{22}$H$_{28}$ with C$_{3v}$ symmetry and in  C$_{26}$H$_{30}$ the difference is as large as the actual renormalization (100\%, 181 vs 363 and 159 vs 316 meV, respectively).\\

\begin{figure}[hb!]
\includegraphics[width=3.4in]{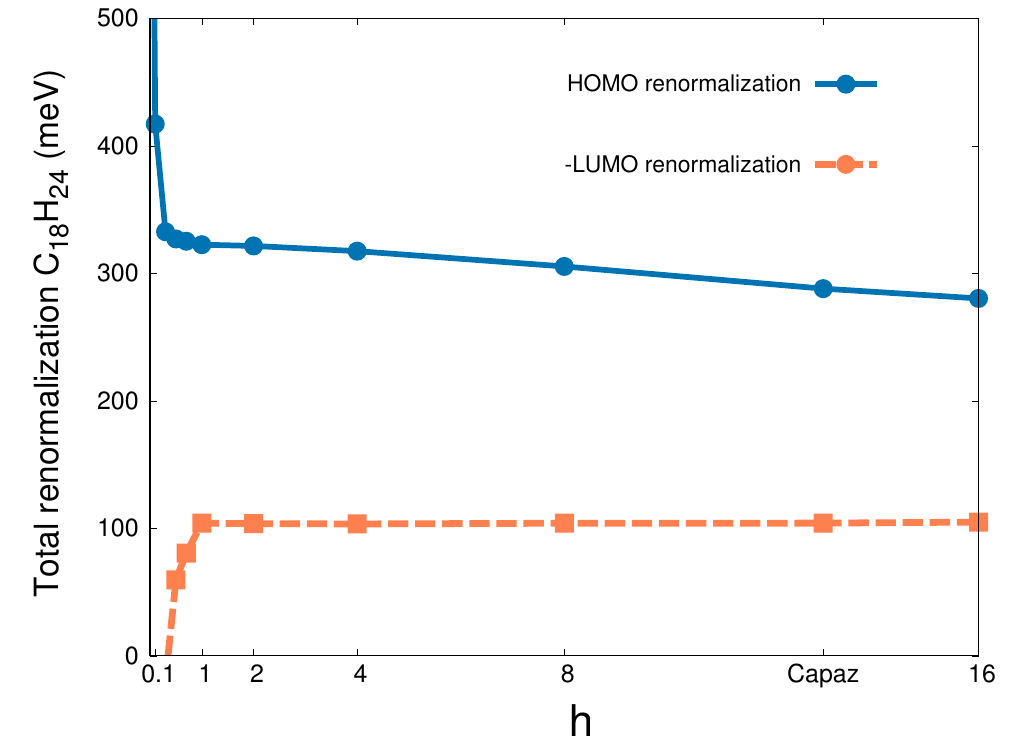}
\includegraphics[width=3.4in]{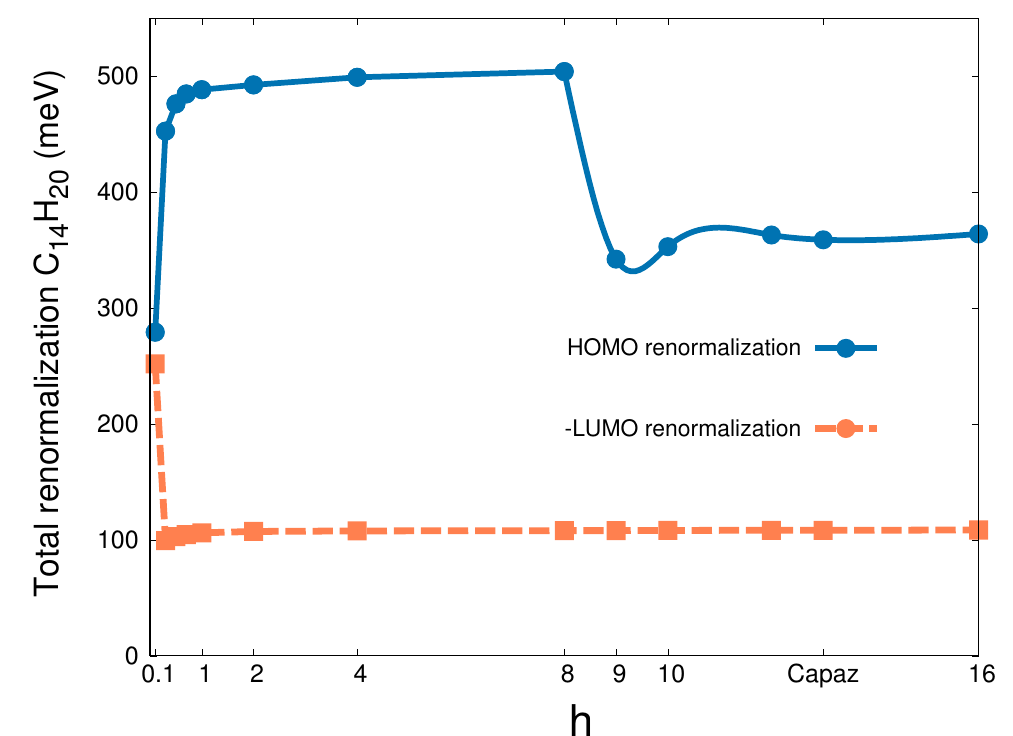}
\caption{Total renormalization (considering all vibrational modes) of HOMO and LUMO (plotted with opposite sign) of C$_{18}$H$_{24}$ (top) and C$_{14}$H$_{20}$ (bottom) as a function of the size of the displacement.}
\label{fig:Renorm_vs_h_C18_C14}
\end{figure}

In Fig. \ref{fig:Renorm_vs_h_C18_C14}, we present the energy renormalization of the HOMO and the LUMO states of diamantane (C$_{14}$H$_{20}$) and triamantane (C$_{18}$H$_{24}$) as a function of the size of the finite-difference displacement $h$. Result obtained by using the original displacement (Eq.(\ref{eqcoefscapaz})) are also included. In Fig. \ref{fig:Renorm_vs_h_C18_C14} we notice that there are two inaccurate regions: those of too small $h$ and of too large $h$. If $h$ is too small, then the displaced nuclear configuration is too similar to the undisplaced one, and the numerical errors become large. If $h$ is too large, then two problems appear: crossover of states and appearance of high-order contributions. 

The HOMO renormalization (blue line) in the top panel of Fig. \ref{fig:Renorm_vs_h_C18_C14} shows a drift with the size of the displacements ($h$). We conclude that $h$ must be high enough to avoid numerical noise but as low as possible to avoid drifts as the one displayed in Fig. \ref{fig:Renorm_vs_h_C18_C14}-top. The mixing of states can be checked by calculating the overlap between undisplaced and displaced wavefunctions $\chi \equiv | \langle \textrm{u} | \textrm{d} \rangle |^2$. 

The mixing of states of certain vibrational modes are given in Tables \ref{tab.1} and \ref{tab.2}  and in Figure \ref{fig:Renorm_vs_Overlap}. In Tables \ref{tab.1} and \ref{tab.2}, we present the squared overlap of displaced and undisplaced wavefunctions $\chi$ averaged for all involved vibrational modes, 
the number of displacements where $\chi$ is lower than 0.95,  
and the averaged squared overlap as a function of the displacement amplitude $h$. In the tables we notice that low displacements lead to high averaged overlaps, but higher displacements lead to increasing state mixing.

\begin{figure}[ht!]
\includegraphics[width=3.4in]{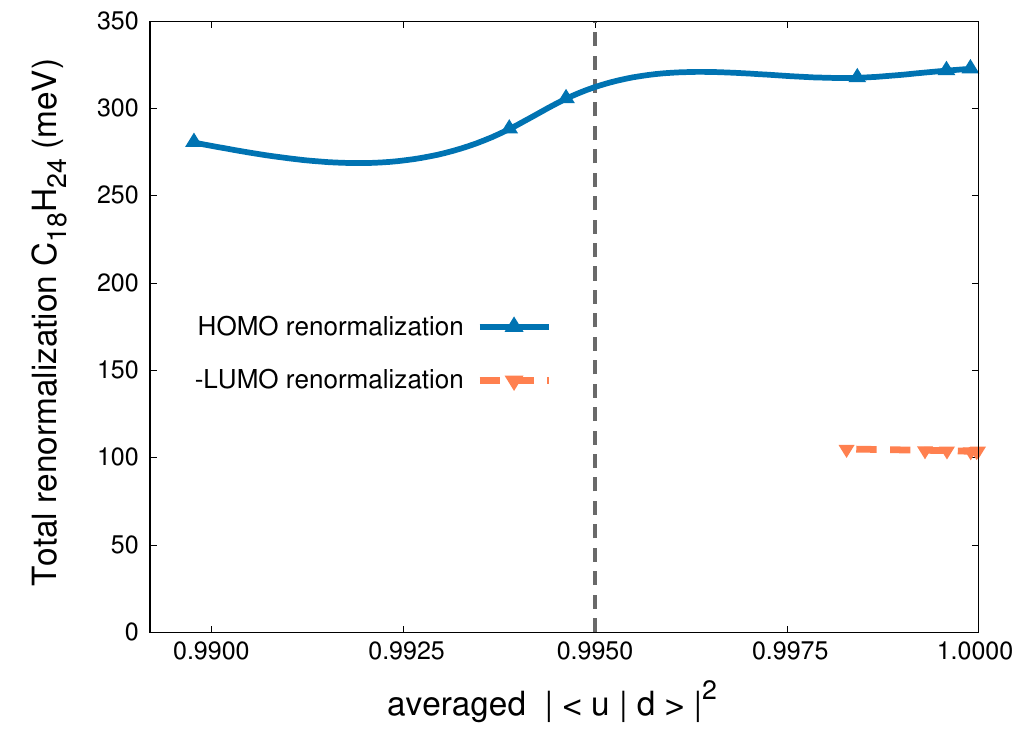}
\includegraphics[width=3.4in]{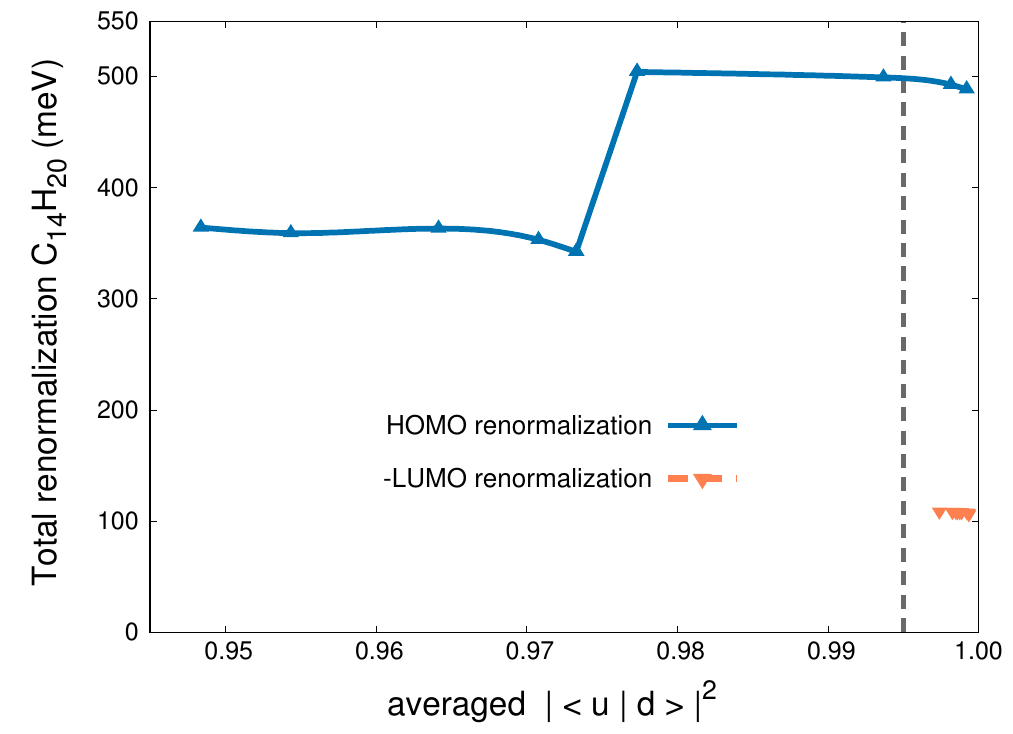}
\includegraphics[width=3.4in]{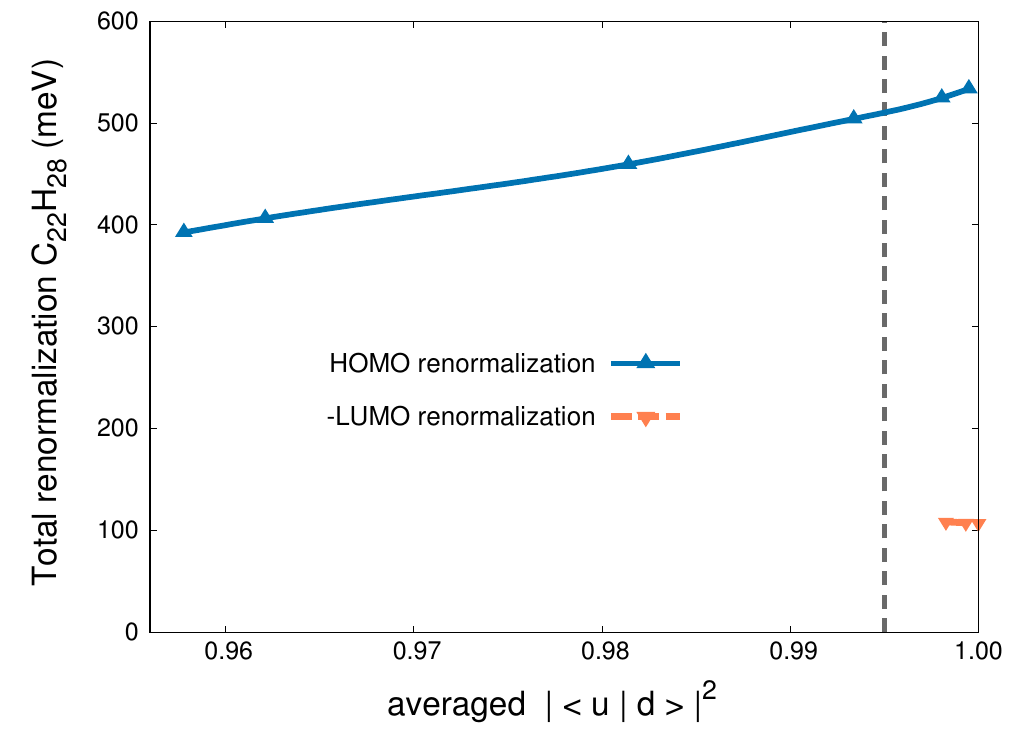}
\caption{Renormalization of HOMO and LUMO (shown with opposite sign) of three diamondois as a function of the averaged overlap between undisplaced and displaced positions
($\overline{ \chi}  \equiv \overline{ | \langle \textrm{u}  |  \textrm{d}  \rangle |^2 }$); each point corresponds to a given value of $h$.
}
\label{fig:Renorm_vs_Overlap}
\end{figure}

\begin{table}
\begin{center}
\begin{tabular}{lclcclrlc|}
\hline
     &	$\overline{\chi } \equiv $                                          && Num. modes 	                        & $\overline{\chi } $  for modes \\
$h$	&	$\overline{|\langle \textrm{u} | \textrm{d} \rangle|^2 }$ 	&& with $\overline{\chi }  < 0.95$ 	&  with  $\overline{\chi }  < 0.95$ \\
\hline
1		&	0.9999		&&	0		&	-\\
2		&		0.9996	&&		0		&	-\\
4		&		0.9984 	&&		2		&	0.9457\\
8		&		0.9946 	&&		4		&	0.8704 \\
{\scriptsize Original}		&		0.9939	&&	3		&		0.8160 \\
16		&		0.9842		&&	16		&	0.8643\\
\hline
\end{tabular}
\caption{Averaged (for all modes) overlaps between undisplaced $| \textrm{u} \rangle $ and displaced $| \textrm{d} \rangle$ HOMO's of C$_{18}$H$_{24}$. ``Original'' refers to the original approach from (Eq.(\ref{eqcoefscapaz})).}
\label{tab.1}
\end{center}
\end{table}

\begin{table}
\begin{center}
\begin{tabular}{lcccc|}
\hline
     &	$\overline{\chi } \equiv $                                          & Num. modes 	                        & $\overline{\chi } $  for modes \\
$h$	&	$\overline{|\langle \textrm{u} | \textrm{d} \rangle|^2 }$ 	& w/ $(\chi  < 0.95)$ 	&  w/  $(\chi  < 0.95)$ \\
\hline
1	&	0.9992	&	0	&	-\\
2	&	0.9982	&	0	&	-\\
4	&	0.9937	&	5	&	0.9079\\
8	&	0.9773	&	19	&	0.8314\\
9	&	0.9733     &	24	&	0.8337 \\
10	&	0.9708	&	27	&	0.8371\\
12	&	0.9642	&	33	&	0.8308\\
{\scriptsize Original}	&	0.9543 &	46	&	0.8343\\
16	&	0.9484	&	45&		0.8125\\
\hline
\end{tabular}
\caption{Same as Tab. \ref{tab.1} but for C$_{14}$H$_{20}$.}
\label{tab.2}
\end{center}
\end{table}

In conclusion, if we aim to avoid a significative error in the frozen-phonon calculation we
suggest to perform displacements with low $h$ (e.g. $h=2$) and, for the chosen displacement size, to make sure that for all vibrational modes the
overlap between undisplaced and displaced wavefunctions are above 0.995. If this is not the case, we recommend to correct the contribution of such \emph{problematic mode} by using the means presented in the main paper.\\

Let us note that, for our calculations, we read the electronic eigenvalues (in atomic units) with 16 digits, which is essential for accurate calculations. 
In addition, in the calculation of the displaced eigenvalues we use the electronic atomic configuration as starting guess (\texttt{startingwfc='atomic'} in \texttt{Quantum Espresso}), for otherwise, if random components appear, the difference with the undisplaced eigenvalues could have undesired different offsets. 
We have performed our calculations in a supercomputing facility (cluster) \cite{notehummel} 
with nodes consisting of two Intel Xeon E5-2630v3 processors, each with 8 cores and with QDR-Infiniband network \cite{GR2012IJMPC}. Our calculations were in general not very much time-consuming for LDA and GGA functionals, but they were more expensive for hybrid functionals. For example, the calculation corresponding to a single mode (out of 168) of C$_{26}$H$_{32}-T_d$ in one node takes about 10 minutes for LDA and GGA functionals (with a cutoff of 30 Ry); if the functional is B3LYP (with a cutoff of 60 Ry), the calculation takes about 4.5 hours in 4 nodes. These timings, however, could be reduced by reducing the size of the simulation box. \\ 

 {
\begin{figure}[h!]
\begin{center}
\includegraphics[width=3.2in]{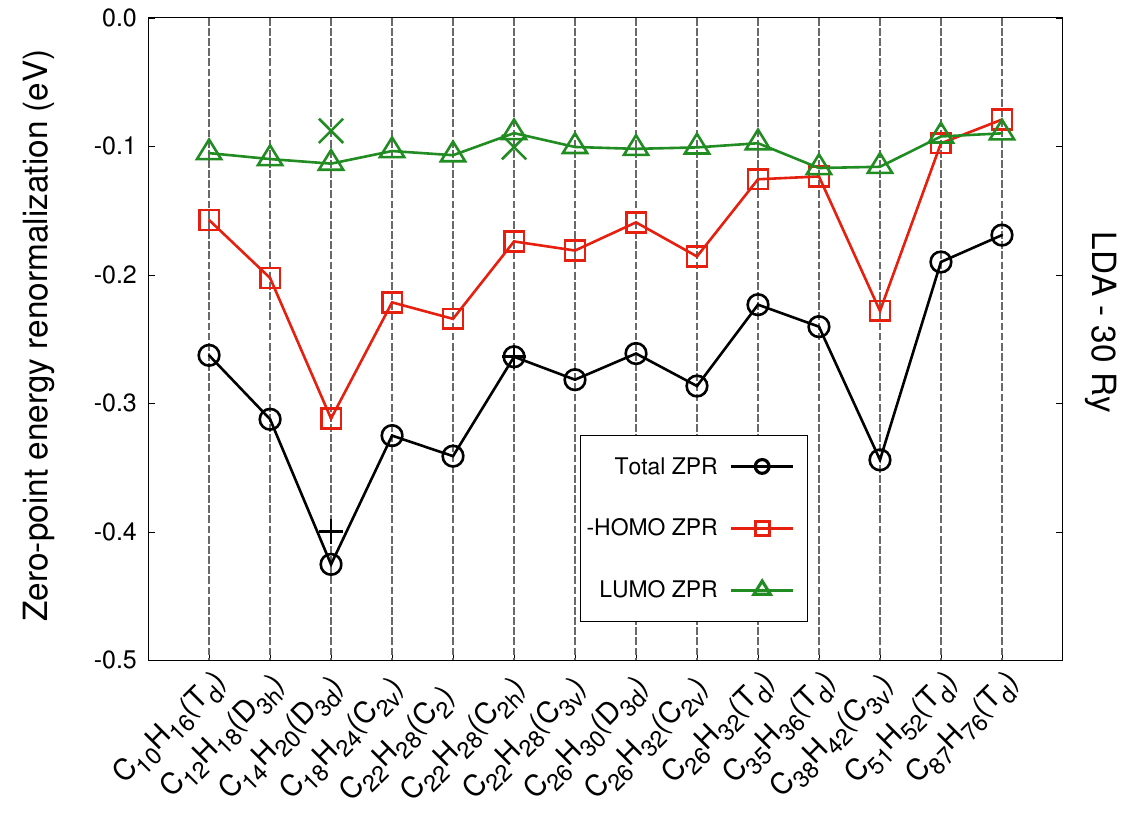}
\includegraphics[width=3.2in]{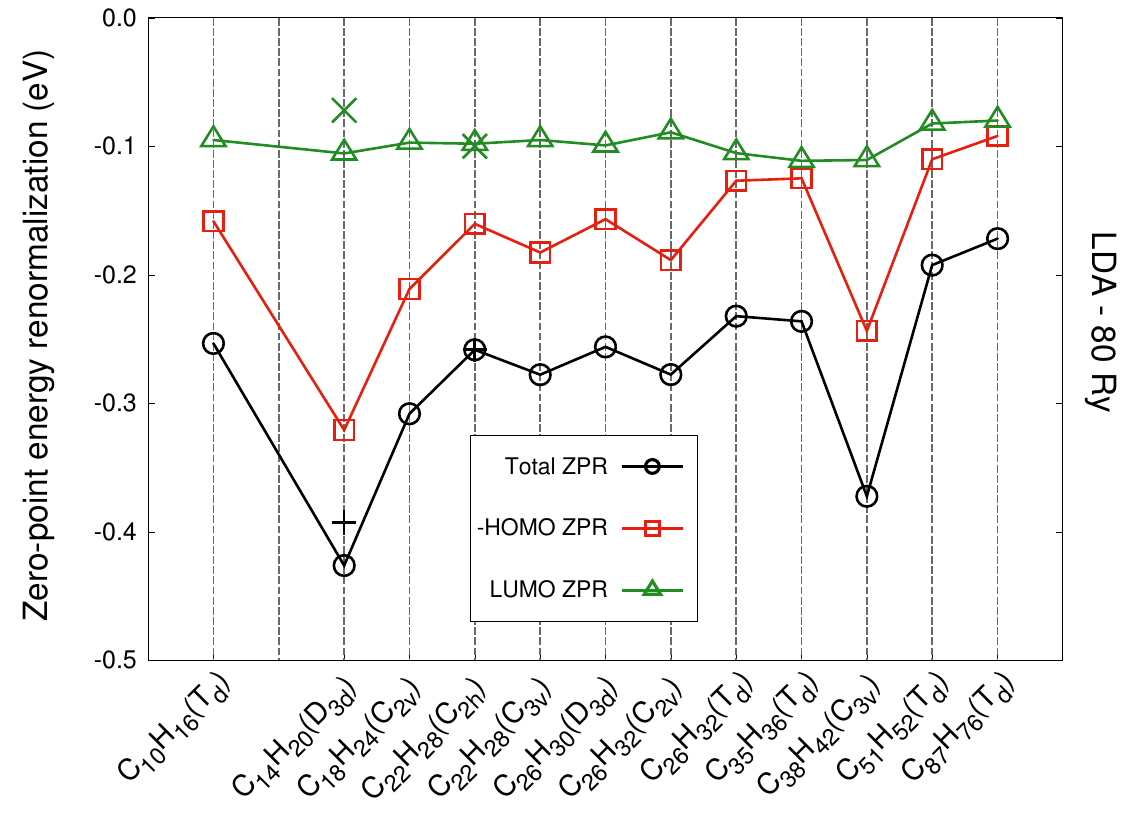}
\includegraphics[width=3.2in]{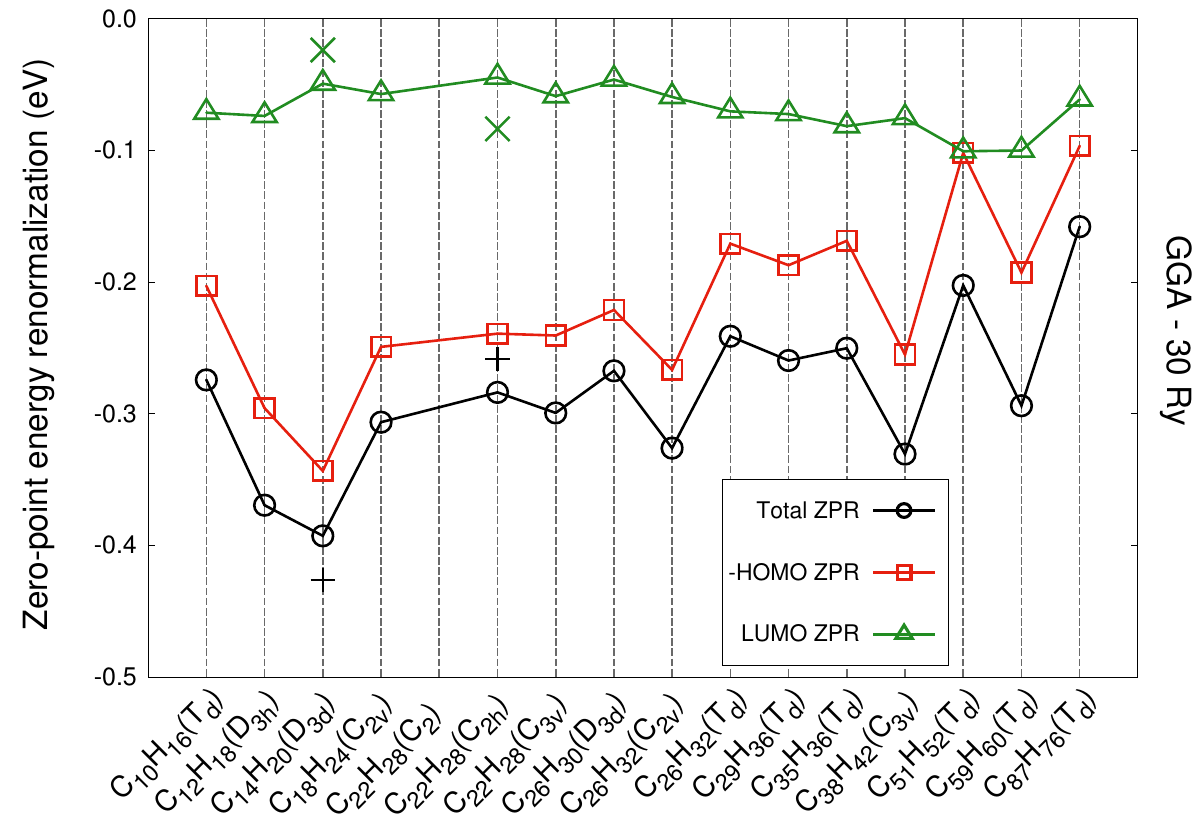}
\includegraphics[width=3.2in]{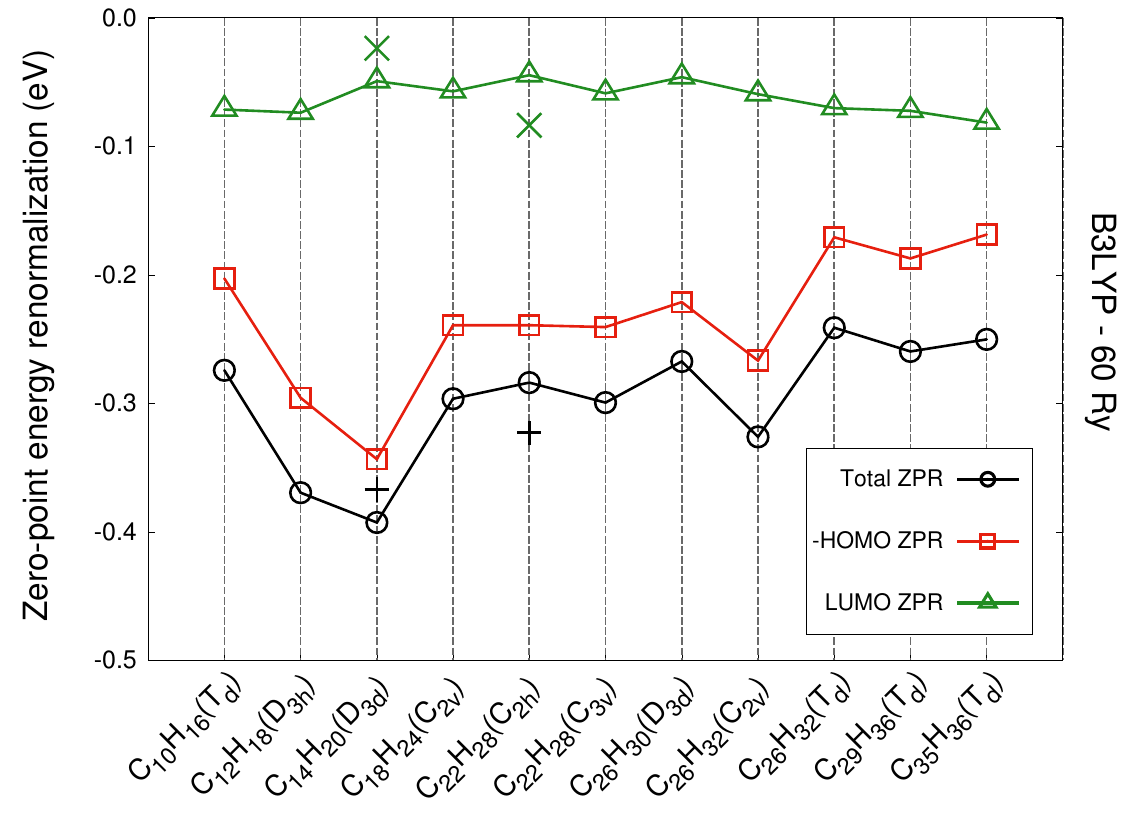}
\caption{\label{Fig.results} {\footnotesize
Zero-point renormalizations of diamondoids using LDA, GGA and hybrid functionals. Top, left: LDA-PZ functional with cutoff of 30 Ry; Top, right: LDA-PZ functional with cutoff of 80 Ry; Bottom, left: GGA-PBE functional with cutoff of 30 Ry; Bottom, right: B3LYP functional with cutoff of 60 Ry (using the HGH pseudopotential with the BLYP functional; the relaxed geometries and normal modes were calculated with the PBE functional with HGH pseudopotential and a cutoff of 30 Ry).
The $\times$ sign indicates renormalization of the LUMO+1, and the $+$ sign is the HOMO/LUMO+1 gap.
}}
\end{center}
\end{figure}
}

\begin{figure}[h!]
\begin{center}
\includegraphics[width=3.2in]{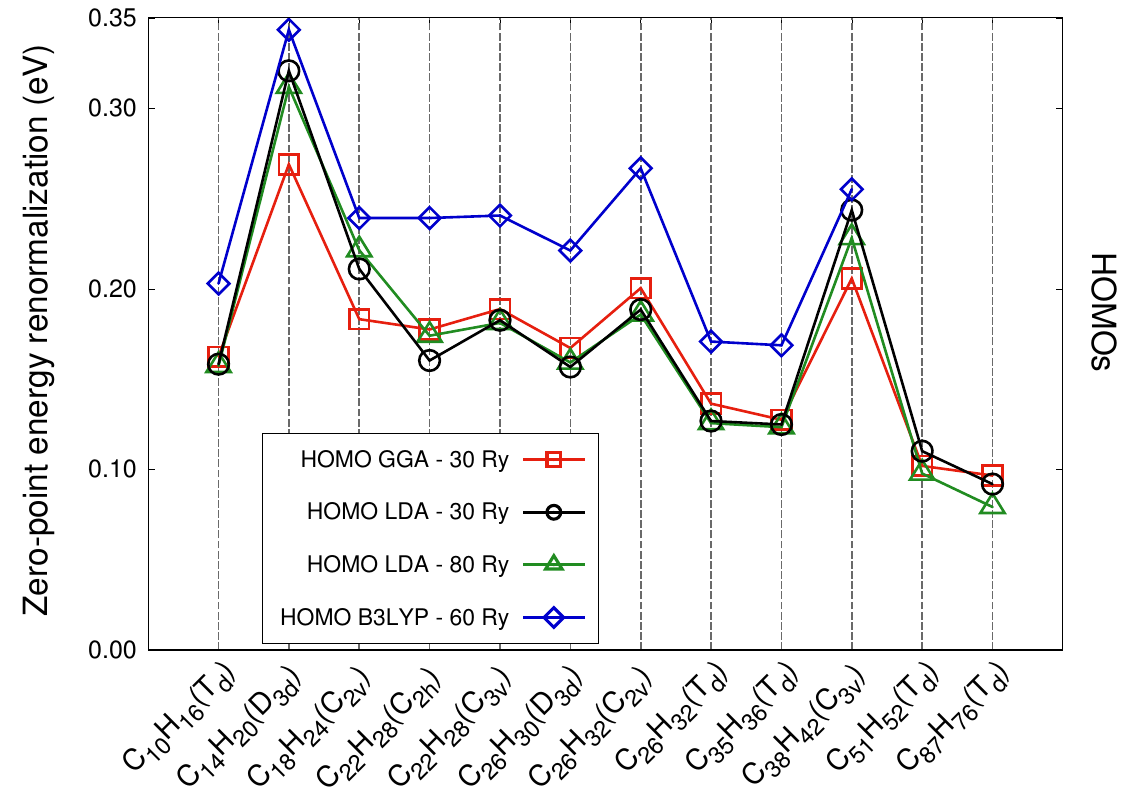}
\includegraphics[width=3.2in]{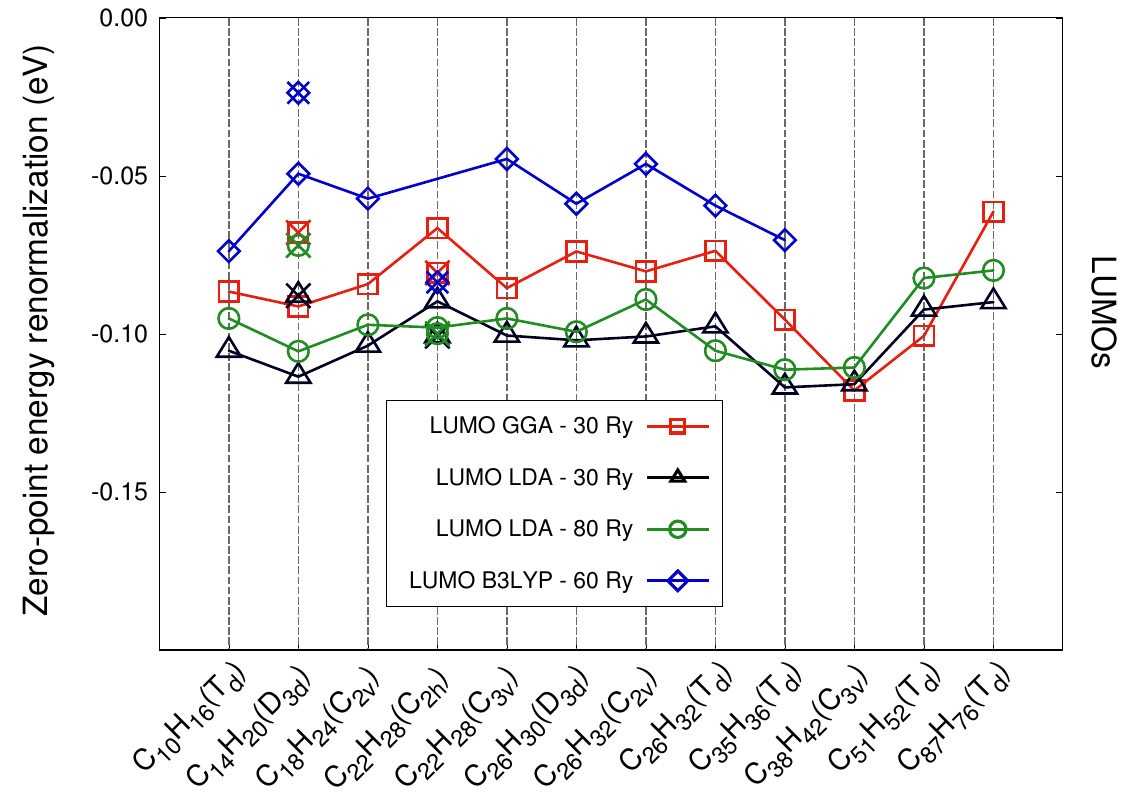}
\includegraphics[width=3.2in]{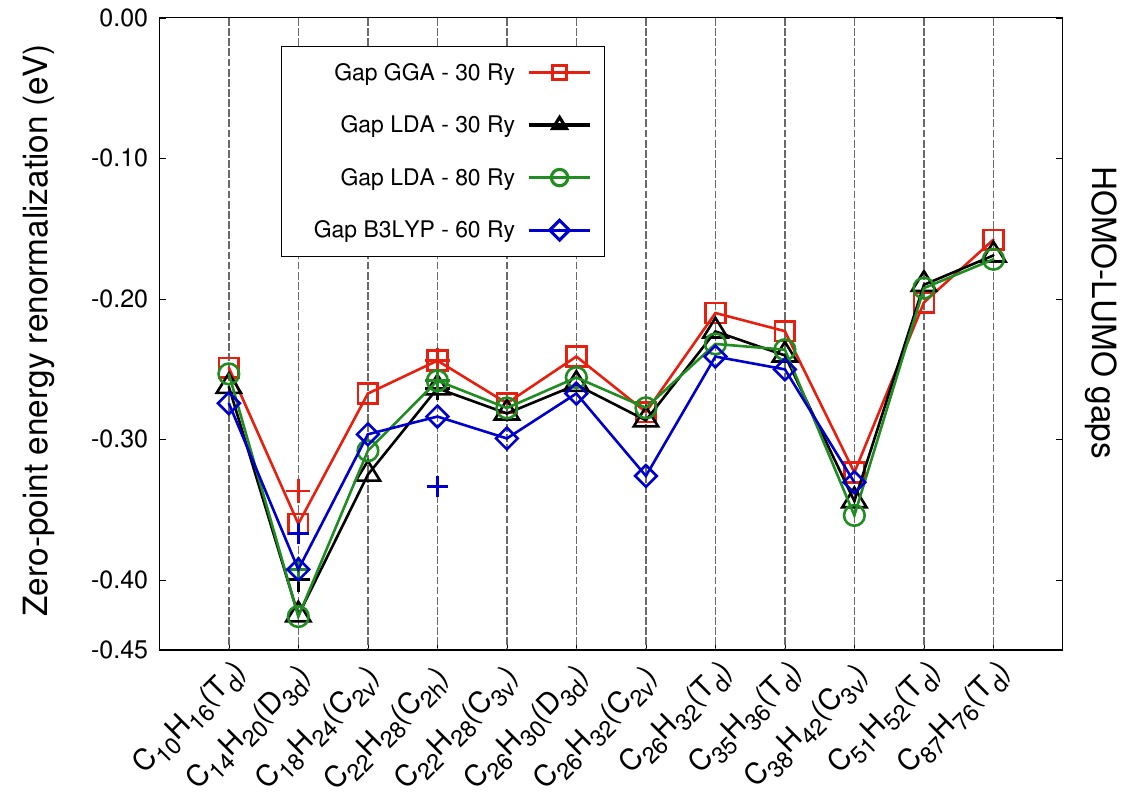}
\caption{\label{Fig.hlg} {\footnotesize
Zero-point renormalizations of diamondoids.
Top Left: HOMO renormalizations;
Top Right: LUMO renormalizations;
Bottom: HOMO-LUMO gap renormalizations.
In the top right graph the points with an $\times$ inside indicate renormalization of the LUMO+1; in the bottom graph the $+$ signs indicate HOMO/LUMO+1 gap.
}}
\end{center}
\end{figure}

 \newpage 

 {
\begin{table}[ht!]
{\scriptsize
\begin{tabular}{ccccccccccccc}
    \hline
    \hline
    \multicolumn{2}{c}{RENORMALIZATIONS}{}&        GGA   &            &       &    LDA          &        &              &         LDA     &              &           &         B3LYP     &     \\
    \multicolumn{2}{c}{(meV)}{}	 	               &         &		 &   	 &     30 Ry       &        &                &     80 Ry     &              &          &                &     \\
\hline
\hline
							&  HOMO	&	162.4   &	            &           &  157.3 &      &    &   158.3       &          &           &  202.9     &    \\
	C$_{ 10}$H$_{16 } - T_d$		& LUMO	&	-86.6   &		    &   	 &   -105.3 &      &    &     -95.1      &        &             &   -71.4    &    \\
							& Gap	&	-249.0   &		    &   	 & -262.6  &      &    &      -253.4      &       &              &  -274.3     &   \\
\hline	
							&  HOMO	& 224.2	  &		    &   	 & 202.5  &	    &   		 &         &   &              &  295.7      &   \\
	C$_{ 12}$H$_{18} - D_{3h}$	& LUMO	& -92.2	   &		    &   	 &  -109.9   &	    &   		 &         &   &              & -73.8    &    \\
							& Gap	&-316.4	   &		    &   	 &  -312.4  &	    &   		 &        &    &              & -369.5     &    \\
\hline
							&  HOMO	        &	268.9 &	{\bf  8}     &   		 &  322.2   & {\bf  6}       &   &    317.7    &    {\bf  4}         &         &   341.4      & {\bf  8}  \\ 
	C$_{ 14}$H$_{ 20} - D_{3d}$	& LUMO	        &	-91.3   &		    &   	 &  -113.5  &      &    &    -105.5     &          &           &   -49.3     &   \\
	                                                  & LUMO+1	&	-67.9	   &		    &   	 &  -87.9  &       &   &      -72.0     &             &        & -23.7    &     \\
							& Gap (H/L)	&	-360.2   &		    &   	 &  -435.7  &      &    &     -423.2    &           &          &    -390.7   &   \\
							& Gap (H/L+1)	&	-336.8	    &	    &   	 &    -410.1     &         &  &    -389.7    &          &           &   -365.1      &     \\
\hline	
							&  HOMO	&	183.2  &	{\bf  2} 	    &   &  221.4  &    {\bf  2}    &   &  211.0      &       {\bf  2}       &        &   239.2    &  {\bf  4}   \\
	C$_{18}$H$_{24} - C_{2v}$	& LUMO	&	 -84.2  &		           &   	 &  -103.7 &      &    &           -97.1     &           		 &         &   -57.2     &   \\
							& Gap	&	-267.4  &		          &   	 &  -325.1  &      &    &        -308.1      &            &      			   & -296.4     &   \\
\hline	
							&  HOMO	&    &		    &   	 &  234.2  &  {\bf  18}     &      &       &              &      &          &   \\
C$_{ 22}$H$_{ 28} - C_{2}$	 	& LUMO	&	 &		    &   	 &  -106.9  &   &     &        &              &       &      &   \\
							& Gap	&	  &		    &   	 &  -341.1  &   &      &         &              &      &       &   \\
\hline	
							&  HOMO	& 177.5	     &  {\bf  1}    &		&   174.0      &  {\bf  2}  &      &     160.3      &     {\bf  1}    &             &  239.2     &  {\bf  2}   \\
C$_{ 22}$H$_{ 28} - C_{2h}$	 	& LUMO	& -66.5	     &     &			         &  -89.7        &   &      &   -98.0        &          &        &   -44,6     &   \\
							& LUMO+1 	&	-80.7	      &    &			 &  -100.7     &     &     &      -99.9     &         &        &   -83.6     &      \\
							& Gap (H/L)	& -244.0	      &    &			 & -263.7     &     &     &   -258.3    &          &          &  -283.8     &   \\
							& Gap (H/L+1)	& -258.2   &     &			 &    -274.7  &     &     &     -260.2     &         &            &  -322.8     &     \\
\hline	
							&  HOMO	&	188.8   &		    &   	 &  181.1  &   &      &    182.6       &              &      &   240.6      &   \\
C$_{ 22}$H$_{ 28} - C_{3v}$	 	& LUMO	&	-85.5   &		    &   	 &  -100.5  &   &     &      -95.1 *     &              &       &   -58.8     &   \\
							& Gap	&	-274.3   &		    &   	 &  -281.6  &   &      &      -277.7     &              &      &   -299.4      &   \\
\hline	
							&  HOMO	&	167.2  &	    &   	 &   159.2&   &      &     156.7      &             &        &  221.2     &    \\
	C$_{26}$H$_{30} - C_{3d}$	& LUMO	&	-73.9  &	    &   		 &   -102.0 &   &     &       -99.3     &         &            &   -46.2     &   \\
							& Gap	&	 -241.1  &	    &   	 &   -261.2 &   &       &     -256.0     &              &       &  -267.4     &    \\							
\hline	
							&  HOMO	&	200.4  &	    &   		 &  185.7 &   &      &     188.5      &           &          &   266.7    &    \\
C$_{ 26}$H$_{ 32} - C_{2v}$			& LUMO	&	-80.2   &    &   			 & -100.8       &   &   &   -89.1  *       &     &              & -59.4     &     \\
							& Gap	&	-280.6   &		    &   	 & -286.5  &   &      &      -277.6     &         &            & -326.1      &    \\							
\hline	
							&  HOMO	& 136.4	   &		    &   	 & 125.6  &       &   &  126.8         &         &            &  170.8      &   \\
C$_{ 26}$H$_{ 32} - T_d$			& LUMO	& -73.7 * &		    &   	 &  -97.6  &       &   &   -105.3      &          &           &   -70.3 *      &   \\
							& Gap	& -210.1	   &		    &   	 &  -223.2 &       &   &   -232.1      &         &            &   -241.1     &    \\
							&  HOMO	& 175.2	   &		    &       &       &   	 &   &   &          &              &  187.2     &   \\
C$_{ 29}$H$_{ 36} - T_d   $			& LUMO	& -93.9 * &	    &       &       &   		 &     &   &         &              &   -72.4     &    \\
							& Gap	& -269.1	   &		    &       &       &   	 &    &   &        &              &  -259.6    &      \\
\hline	
							&  HOMO	&   127.5   &	    &   		 & 123.4  &       &   &    124.9      &          &           &  168.7      &   \\
C$_{35}$H$_{36}  - T_d  $		& LUMO	& -95.5 *   &	    &   		 &   -116.9   &     &      &   -111.3       &       &              & -81.5       &   \\
							& Gap	&  -223.0	   &	    &   		 &   -240.3 &      &    &    -236.2       &        &             & -250.2      &   \\
\hline
							&  HOMO	&  205.7	   &	{\bf  4} 	    &   & 228.1  & {\bf  4}   &        &    243.6     &    {\bf  4}            &      &  255.1   &    {\bf  4}        \\
C$_{38}$H$_{42}  - C_{3v}  $		& LUMO	& -117.9    &		    &   	 &   -115.9  &   &       &    -110.6       &          &               &   -75.5 * &   \\
							& Gap	&  -323.6	   &		    &   	 &   -344.0  &   &       &    -354.2      &           &              &   -330.6 &    \\
\hline
							&  HOMO	&  101.9	   &		    &   	 & 97.7  &     &     &    110.1      &          &       &           &    \\
C$_{51}$H$_{52}  - T_d  $		& LUMO	&  -100.6   &		    &   	 &  -92.3   &      &    &     -82.3     &        &       &             &    \\
							& Gap	&  -202.5	   &		    &   	 &  -190.0  &     &     &    -192.4     &       &       &              &     \\
\hline
							&  HOMO	&   192.8	   &		    &   	 &       &       &       &    &   &        &              &     \\
C$_{59}$H$_{60}  - T_d  $		& LUMO	&  -101.1 *   &		    &   	 &        &       &       &    &   &        &              &     \\
							& Gap	&   -293.9	   &		    &   	 &        &       &       &    &   &        &              &      \\
\hline
							&  HOMO	&  96.6	&    &   &    79.2 &   &       &    91.9        &             &       &        &  \\
C$_{87}$H$_{76}  - T_d  $		& LUMO	&  -61.3 **  &	    &   		 &  -89.9   &       &   &   -79.8      &       &       &              &     \\
							& Gap	&  -157.9   &	    &   		 &  -169.1  &      &    &  -171.8        &       &       &              &    \\
\hline
\hline
							&  HOMO	& -29.5	  &		    &   	 &  -35.4  &		    &       &   	 &      &              & -11.9      &    \\
C$_{ 6}$H$_{12}$ N$_{ 4}  - T_d   $	& LUMO	&  -68.7    &		    &   	 &   -82.3	   &		    &       &   	 &     &              &  -50.3    &    \\
							& Gap	&  -39.2   &		    &   	 &   -46.9   &		    &       &   	 &    &              &   -38.4	    &   	 \\					
\hline
\hline
\end{tabular}
}
\caption{Zero-point renormalizations of the HOMOs, LUMOs and HOMO-LUMO gaps of diamondoids. The numbers in bold font indicate the number of modes where a correction due to an anticrossing was necessary.
{$ \quad$}
} \label{tab:all_results}
\end{table}
}

\newpage


\section*{Comparison of the results of different approaches}\label{sec:comp}

In Table \ref{tab:all_results}, Figures \ref{Fig.results} and \ref{Fig.hlg} we present the renormalizations of HOMOs, LUMOs and gaps for the 4 different kind of calculations: using the GGA-PBE functional with a cutoff of 30 Ry, using the LDA-PZ functional with 30 Ry and 80 Ry, and using B3LYP with a cutoff of 60 Ry.
All calculations of Tables \ref{tab:all_results} were performed with $h=2$, except those marked with * (converged with $h=6$), ** (converged with $h=10$) and the LUMO+1's (converged with $h=8$).
Our $h$ values are similar to those used in Ref. [\onlinecite{ponceprb}]. The results from LDA for cutoffs of 30 Ry and 80 Ry are nearly identical. The zero-point renormalization is thus converged with 30 Ry.

\section*{Electronic states and wavefunctions of adamantane and urotropine}

The electronic state energies along with the HOMO state wavefunctions of adamantane (C$_{10}$H$_{16}$) and urotropine (C$_6$N$_4$H$_{12}$) obtained from DFT-LDA calculations with an energy cutoff of 30~Ry are presented in Fig.~\ref{fig:electronicwfc}(a)-(d). Comparing Fig.~\ref{fig:electronicwfc}(c) and (d), we see the HOMO state wavefunction of urotropine is constituted of the lone pairs and delocalized mainly outside the molecule while that of adamantane is distributed along the carbon-carbon and carbon-hydrogen bonds and localized inside the ``cage''. The delocalized wavefunctions for both the HOMO and the LUMO (not shown) states of urotropine result in similarly small values of ZPR.

\begin{figure}[h!]
\includegraphics[width=5.4in]{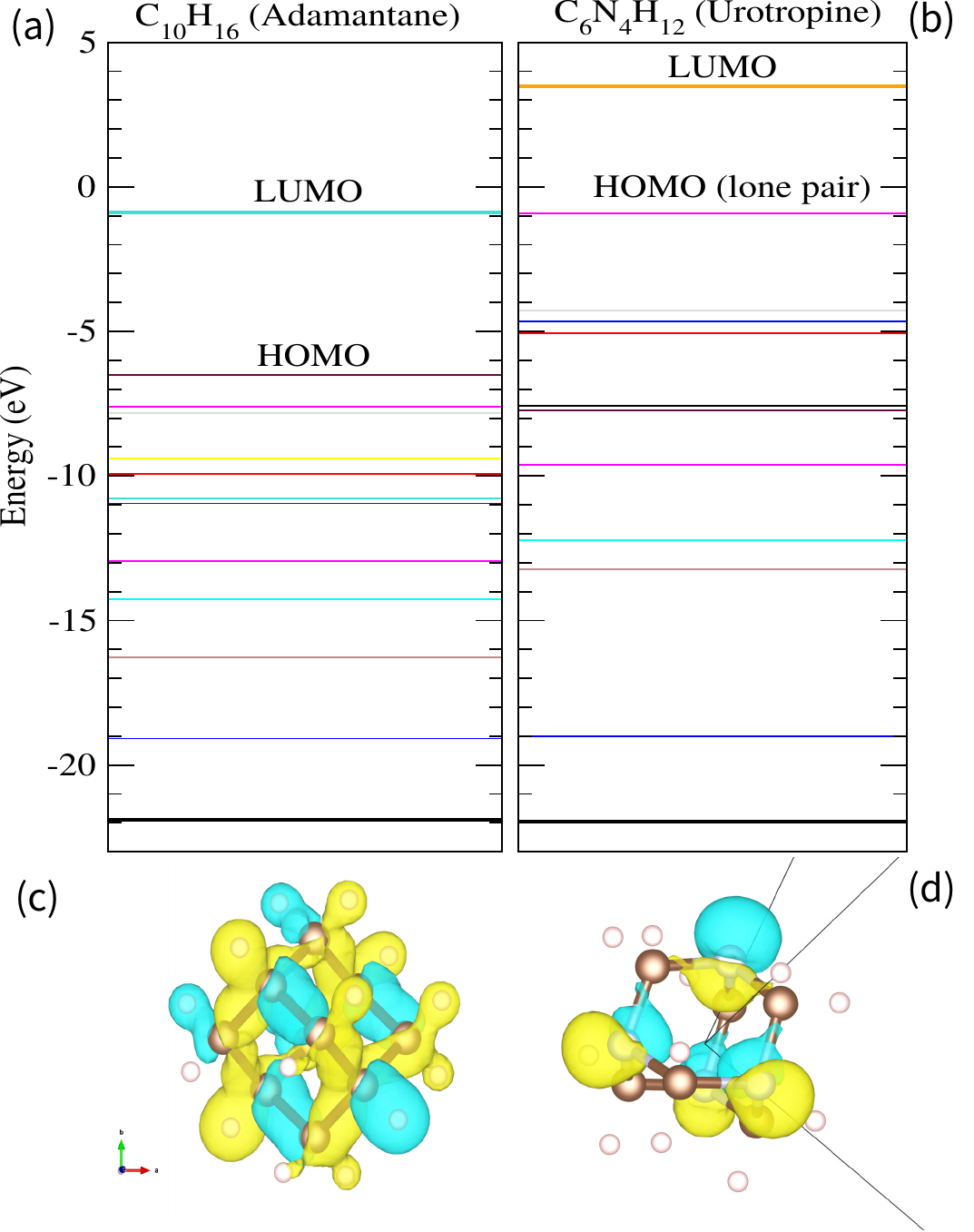}
\caption{The electronic states of (a) adamantane (C$_{10}$H$_{16}$) and (b) urotropine (C$_6$N$_4$H$_{12}$) obtained from DFT-LDA calculations with energy cutoff of 30~Ry and arbitrarily aligned at the lowest eigenvalue. The HOMO state squared wavefunction of (c) adamantane and (d) urotropine.
}\label{fig:electronicwfc}
\end{figure}

\clearpage

\bibliographystyle{unsrt}

 \end{widetext}


%

\end{document}